\documentclass[galaxies,article,accept,moreauthors,pdftex,10pt,a4paper]{mdpi}

\usepackage{amsmath}
\usepackage{amssymb}
\usepackage{aas_macros}

\usepackage{booktabs}
\usepackage{multirow}
\usepackage{soul}
\usepackage{microtype}

\setitemize{parsep=6pt,itemsep=0pt,leftmargin=*,labelsep=5.5mm}
\setenumerate{parsep=6pt,itemsep=0pt,leftmargin=*,labelsep=5.5mm}
\setlist[description]{itemsep=0mm}

\newcommand{\survey}[1]{\textsl{#1}}

\newcommand{\documentname}{\emph{Article}}

\newcommand{\kpc}{\mathrm{kpc}}
\newcommand{\msun}{\mathrm{M}_\odot}
\newcommand{\kms}{\mathrm{km}~\mathrm{s}^{-1}}
\newcommand{\frrmg}{\ensuremath{f_{\rm RR:MG}}}

\newcommand{\feh}{\ensuremath{[{\rm Fe}/{\rm H}]}}
\newcommand{\ion}[2]{#1\,\textsc{#2}}

\firstpage{1}
\articlenumber{x}
\doinum{10.3390/------}
\pubvolume{xx}
\pubyear{2017}
\copyrightyear{2017}
\externaleditor{Academic Editors: Duncan A. Forbes and Ericson D. Lopez}
\history{Received: 1 July 2017; Accepted: 14 August 2017; Published: date}

\Title{Disk Heating, Galactoseismology, and the Formation of Stellar Halos}

\Author{Kathryn V. Johnston
 $^{1,}$*$^{,\dagger}$, Adrian M. Price-Whelan $^{2,\dagger}$, Maria Bergemann $^{3}$, Chervin Laporte $^{1}$, Ting~S.~Li $^{4}$, Allyson A. Sheffield $^{5}$, Steven R. Majewski $^6$, Rachael S. Beaton $^7$, Branimir Sesar $^{3}$ and Sanjib Sharma $^{8}$}
\AuthorNames{Firstname Lastname, Firstname Lastname and Firstname Lastname}

\address{%
$^{1}$\quad Department of Astronomy, Columbia University, 550 W 120th st., New York, NY 10027, USA;  kvj@astro.columbia.edu (K.V.J.); cfl2126@columbia.edu (C.F.L.) \\
$^{2}$ \quad Department of Astrophysical Sciences, Princeton University, 4 Ivy Lane, Princeton, NJ 08544, USA; adrn@astro.princeton.edu \\
$^{3}$ \quad Max Planck Institute for Astronomy, Heidelberg 69117, Germany; bergemann@mpia-hd.mpg.de (M.B.); bsesar@mpia.de (B.S.)\\
$^{4}$ \quad Fermi National Accelerator Laboratory, P. O. Box 500, Batavia, IL 60510, USA; sazabi@neo.tamu.edu\\
$^{5}$ \quad Department of Natural Sciences, LaGuardia Community College, City University of New York, 31-10~Thomson Ave., Long Island City, NY 11101, USA; asheffield@lagcc.cuny.edu \\
$^{6}$ \quad Department of Astronomy, University of Virginia, P.O. Box 400325, Charlottesville, VA 22904, USA; srm4n@virginia.edu \\
$^{7}$ \quad The Carnegie Observatories, 813 Santa Barbara Street, Pasadena, CA 91101, USA; rlb9n@virginia.edu\\
$^{8}$ \quad Sydney Institute for Astronomy, School of Physics, University of Sydney, NSW 2006, Australia; sanjib.sharma@gmail.com 
}

\corres{Correspondence: kvj@astro.columbia.edu; Tel.: +1-212-854-3884}

\firstnote{These authors contributed equally to this work.}
\abstract{
Deep photometric surveys of the Milky Way have revealed diffuse structures
encircling our Galaxy far beyond the ``classical'' limits of the stellar disk.
This paper reviews results from our own and other observational programs, which together suggest that, despite their extreme positions, the stars in these structures were formed in our Galactic disk.
Mounting evidence from recent observations and simulations implies kinematic connections between several of these distinct structures.
This suggests the existence of collective disk oscillations that can plausibly be traced all the way to asymmetries seen in the stellar velocity distribution around the Sun.
There are multiple interesting implications of these findings:
they promise new perspectives on the process of disk heating;
they provide direct evidence for a stellar halo formation mechanism in addition to the accretion and disruption of satellite galaxies; and,
they motivate searches of current and near-future surveys to trace these oscillations across the Galaxy.
Such maps could be used as dynamical diagnostics in the emerging field of ``Galactoseismology'', which promises to model the history of interactions between the Milky Way and its entourage of satellites, as well examine the density of our dark matter halo.
As sensitivity to very low surface brightness features around external galaxies increases, many more examples of such disk oscillations will likely be identified.
Statistical samples of such features not only encode detailed information about
interaction rates and mergers, but also about long sought-after dark matter
halo densities and shapes.
Models for the Milky Way's own Galactoseismic history will therefore serve as a
critical foundation for studying the weak dynamical interactions of galaxies across the
universe.
}


\begin{document}

%%%%%%%%%%%%%%%%%%%%%%%%%%%%%%%%%%%%%%%%%%

%%%%%%%%%%%%%%%%%%%%%%%%%%%%%%%%%%%%%%%%%%
%\setcounter{section}{-1} %% Remove this when starting to work on the template.

\section{Introduction}

Our perspective on the Milky Way presents both unique challenges and unique opportunities within our quest to understand galaxies more generally.
Because we are located inside of our own Galaxy, it is the only galaxy for which we lack a truly global perspective in a single snapshot and, instead, must survey the entire night sky to fully sample its constituents.
%From our position inside, it is the one galaxy in the Universe that we cannot take an image of but rather need to survey the entire sky in order to map its structure.
On the other hand, it is one of the few galaxies that we can presently study by individual stars and %is the only one for which we can hope to make volume-complete maps of full-space positions and velocities for significant numbers of unevolved stars.
it is the only galaxy for which we can make volume-complete samples in both position and velocity space for non-evolved stellar tracers (e.g., via Main Sequence Turnoff Stars, hereafter MSTO).
Present and recent sky surveys have already considerably advanced this effort, and surveys in the near future will deliver massive datasets that will enable detailed studies of stellar structures throughout the Galactic volume.
%move beyond co-ordinates restricted to random projections in a limited number of directions to

% We are in the middle of a renaissance in Milky Way studies, fueled by stellar data sets of sufficient scope in terms of sky coverage and numbers to exploit our perspective as a strength rather than a liability
% %and take full advantage of our proximity
% to create vast catalogues of stellar data.

Emerging in the 1990s, the catalogues that provided the inspiration for  current and future surveys not only mapped global structures in our Galaxy, but also revealed the ubiquity of substructure within it.
These revelations added an unforeseen richness to interpretations of the data sets and encouraged the development of new dynamical tools for studying ongoing interactions and formation histories.
As a few examples:
\begin{itemize}
    \item Astrometric data from the \survey{Hipparcos} mission \cite{esa97} led
        to the discovery of moving groups  in the velocity distribution
        of solar-neighborhood stars \cite{dehnen98}. Some of these likely
        correspond to destroyed star clusters (as expected), while others
        (unexpectedly) have been interpreted as signatures of resonances with
        the Galactic bar \cite{dehnen00}.
    \item Precise, large-area photometry from the Sloan Digital Sky Survey
        (hereafter, SDSS ---\cite{york00,stoughton02,abazajian03}) led to the
        discovery of many ``streams'' of MSTO stars in
        the Galactic stellar halo. These are understood to be the remnants of
        long-dead satellite galaxies and dissolved globular clusters
        \cite{newberg02,belokurov06} and serve as a stunning confirmation that
        our Galaxy has indeed formed hierarchically
       (e.g., \cite{bullock01,bullock05,bell08}).
    \item All-sky, infrared photometry from the Two Micron All Sky Survey
        (hereafter, 2MASS ---\cite{nikolaev00}) enabled M giant stars
        associated with tidal debris from the Sagittarius dwarf galaxy (hereafter Sgr) to be traced around
        the entire sky \cite{majewski03}, offering a new perspective on the
        history of its disruption~\cite{law10}.
\end{itemize}
From these and many other studies using large survey catalogues, it is now clear that the Milky Way is full of kinematic substructure, from the nearby regions of the Galactic disk to the distant stellar~halo.

\begin{figure}[H]
\centering
\includegraphics[width=5 in]{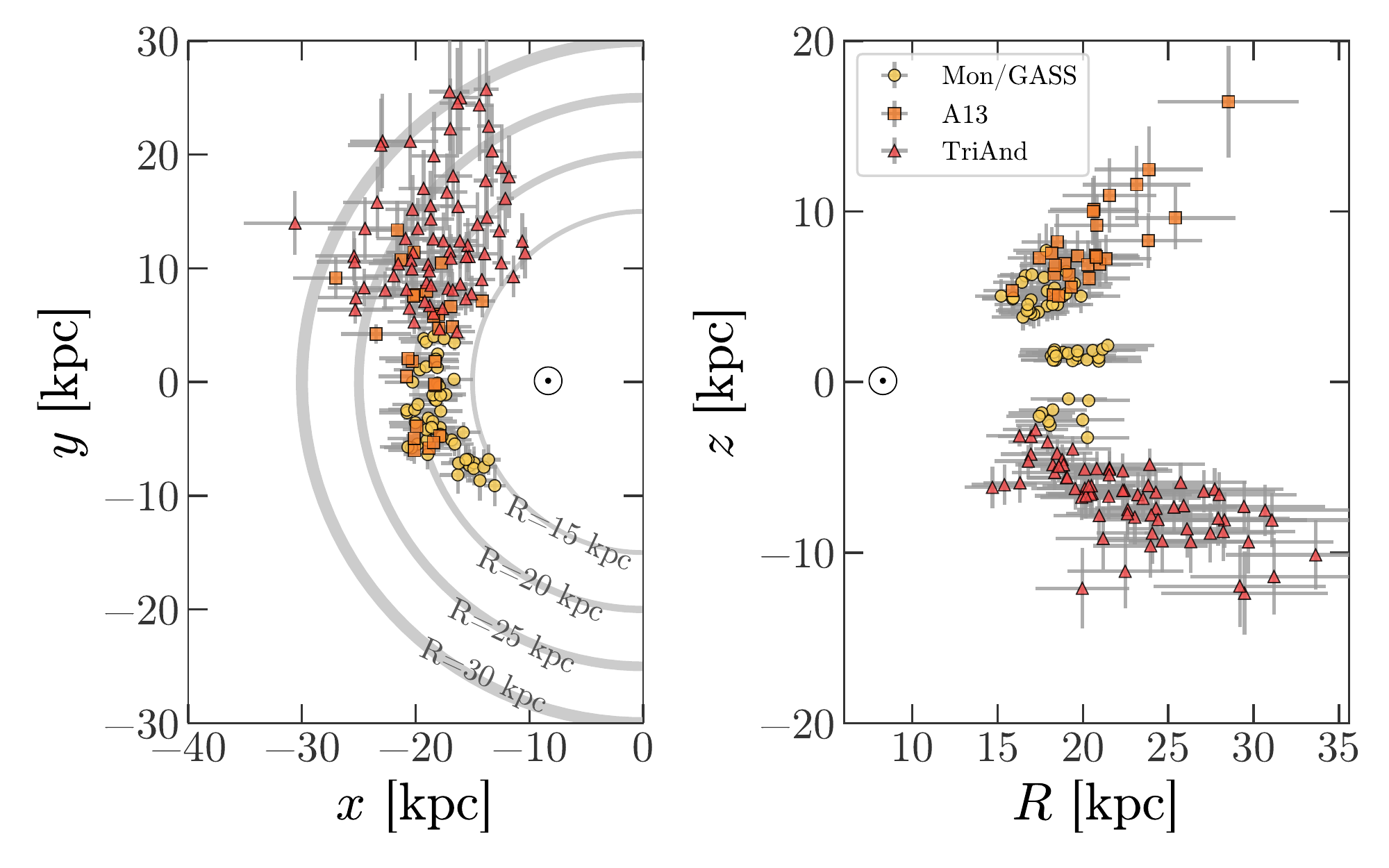}
\vspace{-12pt}
\caption{\label{fig:ting}
Summary of the spatial distribution of M giants in each of the three
low-latitude structures.
Note that at lower Galactic latitudes, the lack of candidate M giant members is
due to selection effects and crowding near the midplane; we expect the
structures to continue towards even lower latitudes, but blend with ordinary
disk stars.
Markers represent individual stars identified as likely members of each of the
three structures discussed in this work (see figure legend).
Distance estimates come from photometry alone and have expected absolute
uncertainties around $\approx$20\% for TriAnd and A13 \cite{sheffield14,li17}
and $\approx$25\% for Mon/GASS.
Grey curves in left panel show Galactocentric circles with cylindrical radii,
$R$, indicated on the figure.
The position of the Sun is marked with the solar symbol $\odot$.
}
\end{figure}

The focus of this \documentname\ is on three such substructures just beyond the
historical ``end'' of the Galactic disk within the inner stellar halo.
Figure~\ref{fig:ting} (reproduced with data from previous work,~\cite{li17}),
shows the spatial distribution of M giant stars associated with these three
substructures: the so-called Monoceros Ring (Mon; also known as the Galactic
Anticenter Stellar Structure, GASS), the Triangulum-Andromeda clouds (TriAnd),
and A13.

Each of these substructures were originally identified as over-densities in
stellar number counts relative to the expected global structure of the disk or
inner stellar halo \footnote{Here we show just M giant stars that have been
previously identified as candidate members of the structures, however, some of
the structures have also been detected in MSTO stars \cite{yanny03,ibata03}.}
The same region of the sky has been shown to be richly structured on even
smaller scales \cite{grillmair06,grillmair08,slater14,martin14,deason14}, but here we consider only the
larger structures.
Unlike most stellar streams, Mon/GASS, TriAnd, and A13 are present at a range
of low to moderate Galactic latitudes and span large areas of the sky---we
will hereafter refer to them collectively as the ``low-latitude structures''.
The basic properties of the low-latitude structures are~summarized~below:
\begin{itemize}
\item{\it{Mon/GASS}} is an arc-like or partial-ring feature of stars beyond the previously-expected edge of the Galactic disk, $\approx 5~\kpc$ beyond the Sun in cylindrical radius \cite{robin92}.
Stars attributed to Mon/GASS span a large area of the sky and large range
in distance: Galactic longitudes $\approx $120$^\circ\lesssim l \lesssim
240^\circ$, latitudes $-30^\circ \lesssim b \lesssim +40^\circ$, and
heliocentric distances $5\lesssim d_\odot \lesssim 10~\kpc$
\cite{yanny03,ibata03,Morganson:2016}.
Radial velocity measurements of M giant stars associated with the structure
follow a clear trend in mean velocity with Galactic longitude and have a
velocity dispersion much smaller than the stellar halo \cite{crane03}.
\item{\it TriAnd} was first discovered as a diffuse over-density of M giant
stars covering the area $\approx$100$^\circ \lesssim~l~\lesssim~160^\circ$,
$\approx $-35$^\circ \lesssim b \lesssim -15^\circ$, overlapping the Mon/GASS
structure on the sky but at larger heliocentric distances of $\approx
$15--25$~\kpc$ \cite{rochapinto04}.
The M giants again exhibit a coherent radial-velocity sequence with a
dispersion much smaller than the halo \cite{rochapinto04}.
Deep photometry in the region revealed MSTO stars associated with the structure
and proposed the existence of a second main sequence (``TriAnd 2'') at larger
distance, $\approx $30--35$~\kpc$ \cite{martin07,martin14}.
\item{\it A13} is another tenuous association of M giants in the North Galactic
Hemisphere in the area $\approx $125$^\circ \lesssim l \lesssim 210^\circ$,
$\approx $20$^\circ \lesssim b \lesssim 40^\circ$ at approximate distances of
$\approx $10--20$~\kpc$.
It was initially discovered by applying a group finding algorithm
\cite{sharma09} to all M giants in the 2MASS photometric catalogue
\cite{sharma10}.
Again, radial velocities of M giants in this structure have a small velocity
dispersion around a roughly linear trend with Galactic longitude \cite{li17}
\end{itemize}

Several distinct scenarios have been used to explain the formation and
existence of each of these structures.
Mon/GASS has been attributed to the accretion of a satellite  \cite{penarrubia05}, a natural extension of the Galactic warp
\cite{momany04,momany06}, or disturbances to the Galactic disk
\cite{kazantzidis08,younger08,purcell11,slater14,xu15,gomez16}.
The extreme position of TriAnd at $(R,Z) \approx (15, -5)$ to $(25,-12)~\kpc$
across a large range in Galactocentric azimuth, $\phi$, seemed to exclude the
possibility of a disturbed Galactic disk as a possible origin and it has also been
modeled as debris from a satellite on a retrograde orbit \cite{sheffield14}.
Initial abundance studies measured $[\alpha/{\rm Fe}]$ and s-process values for M giants in
both Mon/GASS and TriAnd and found them to be consistent with those seen in
Milky Way satellite galaxies, thus unlike the disk \cite{chou2010b,chou11} ---
see Section~\ref{sec:abundances} for a more complete discussion.

Recent evidence points towards a more convincing, coherent picture for the
nature of the three low-latitude structures: the stars in these structures
likely have a common origin in the Galactic disk and have been ``kicked out''
to their present-day positions.
This \documentname\ reviews recent and ongoing contributions that our own group is making  to formulating this picture, which include spectroscopic surveys of the low-latitude structures to study metallicities and kinematic properties \cite{sheffield14,li17}, stellar populations (Sheffield et al., in prep \cite{pricewhelan15}), and detailed abundance patterns (Bergemann et al., in prep.), as well as numerical simulations (Laporte et al., in prep  \cite{sheffield14,pricewhelan15,laporte16}).
We summarize this observational and theoretical work in Sections ~\ref{sec:obs} and~\ref{sec:theory} respectively, adding in the context of contemporary work from other groups, as well as the larger context of possible connections across the Galactic disk.
Armed with this understanding of the nature of these substructures, we proceed in Section ~\ref{sec:Discussion} to discuss prospects for mapping such structures more generally around our own and other galaxies.
We end in Section ~\ref{sec:conc} by outlining the motivation for making such maps, asking what they might be telling us about bigger questions: galaxy formation scenarios and the distribution of dark matter around galaxies.

\section{The Nature of Structures Around the Outer Disk---Summary of Observations}
\label{sec:obs}

From clustering in positions or distance alone, many candidate groups and
over-densities of M giants have been identified in the outer disk or inner halo.
Over the last five years our group has obtained spectroscopy for
candidate members of these structures with the aims of (1) confirming the
existence of substructure in velocities, (2) measuring chemical abundances, and
(3) studying the constituent stellar populations.
These goals then inform our own efforts to produce plausible dynamical
formation scenarios using simulations.
In particular, we {\it avoid} the collimated stellar streams that have been
well-studied in prior work (such as Sgr, Orphan, GD1 and Pal 5---see,
e.g., \cite{law10,koposov10,kuepper15,bovy16}) and instead focus on stellar structures that
appear diffuse, amorphous, extended or  ``cloud-like'' in nature,  such as TriAnd, A13,
and Mon/GASS.
Initial interpretations of these morphologies suggested the structures could be
{\it shells}
---debris from the disruption of satellite galaxies on near-radial
orbits---but as seen from an internal perspective  \cite{johnston08}.
However, our own and other recent observations instead suggest a common origin
within the Galactic disk for stars associated with the low-latitude structures.

\subsection{Low-resolution Spectroscopy: Metallicities and Radial Velocities}

\begin{figure}[H]
\centering
\includegraphics[width=3.5 in]{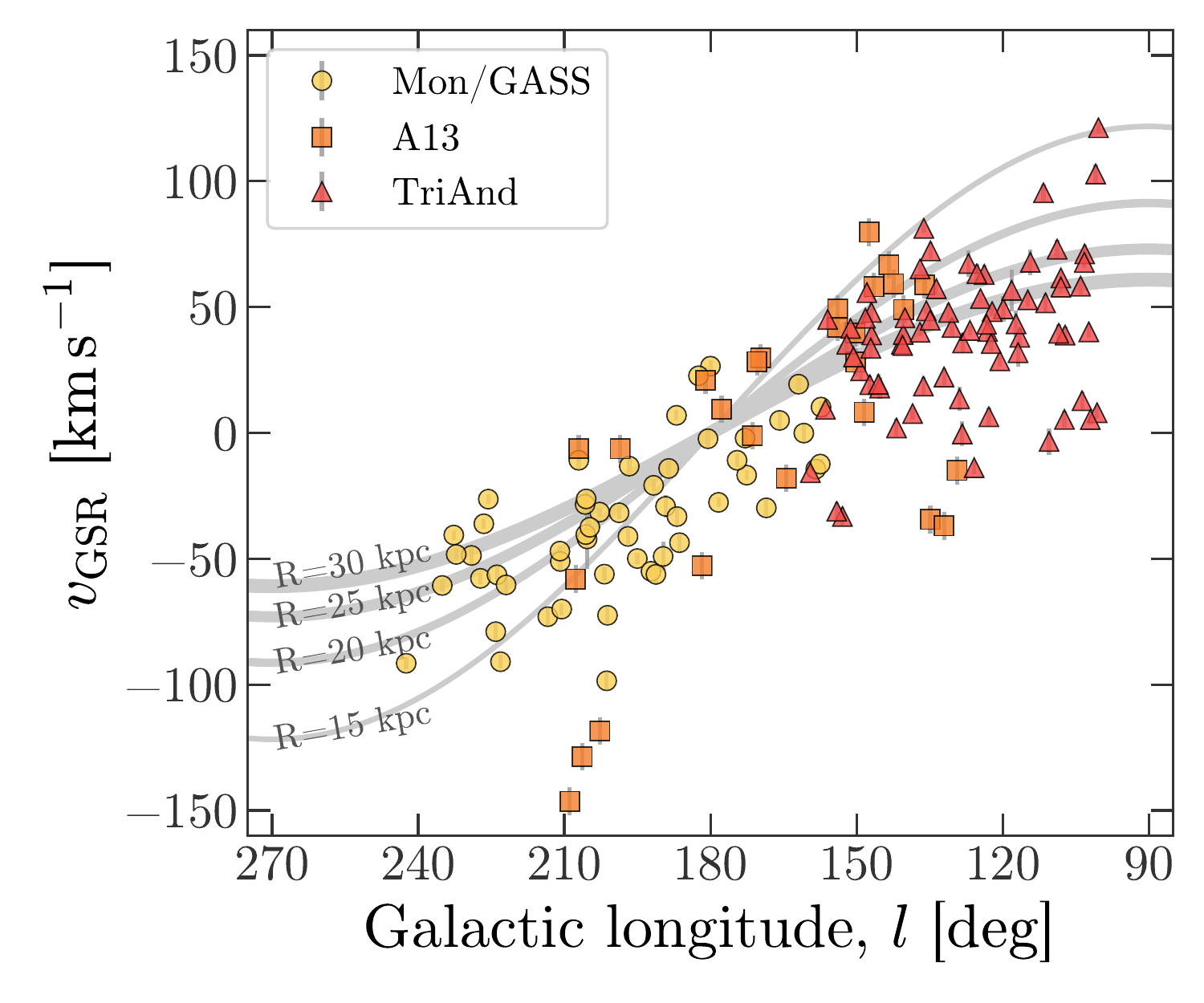}
\vspace{-12pt}
\caption{\label{fig:ting_vel}
Summary of the velocity distribution of M giants in each of the three
low-latitude structures.
Markers represent individual stars identified as likely members of each of the
three structures discussed in this work (see figure legend).
Grey curves show the expected $v_{\rm GSR}$ trends for circular orbits in the Galactic disk midplane with
velocity equal to $220~\kms$ at several Galactocentric cylindrical radii, $R$,
as indicated on the figure.
Velocity uncertainties are typically the same as or smaller than the marker
sizes.
Reproduced from \cite{li17}.
}
\end{figure}

In our first study, we extended a prior sample of TriAnd M giants
\cite{rochapinto04} by obtaining spectra of all candidate M giants identified
by applying color-magnitude cuts to stars in the TriAnd region of the sky
\cite{sheffield14}.~
We~identified M giants associated with the two proposed MSTO TriAnd structures
(TriAnd 1 and 2, as named by \cite{martin07,martin14}).
M giant stars in both TriAnd 1 and 2 form clear over-densities in radial
velocities with a small dispersion, $\sigma_v \approx 25~\kms$, compared to the
background halo velocity distribution.
The radial velocities of M giants in both TriAnd 1 and 2 follow the same
sequence in velocity with a steady negative gradient of mean Galactic
Standard-of-Rest (GSR) radial velocity ($v_{\rm GSR}$) with increasing Galactic
longitude, $l$; see Figures~\ref{fig:ting_vel} and \ref{fig:apw}, red triangles.
We initially presented a dynamical model that simultaneously and approximately
reproduces TriAnd 1 and 2 as tidal debris stripped over two separate pericentric
passages from a single accreted satellite on a low-eccentricity, retrograde,
near-planar orbit.
The debris structures in the simulation were morphologically closer to {\it
streams}  than {\it shells}, but still subtended large areas on the sky
as observed from the Sun's position.
We note, however, that because of large distance uncertainties, the M giants in
TriAnd 1 and 2 are indistinguishable and overlap in distance, velocity
distribution, and sky position; the existence of two distinct structures rather
than a single extended structure has yet to be conclusively demonstrated (see~\cite{martin14} for some counter-arguments).
Hereafter, we therefore refer to the TriAnd structures collectively, rather
than individually.

In subsequent work, we continued this spectroscopic survey by observing M giant
stars in~A13~\cite{sharma10}.
A13 overlaps the TriAnd clouds in Galactic longitude (but not latitude) at one end, and Mon/GASS at
the other end, but is apparent in the Northern (rather than Southern) Galactic Hemisphere and at slightly brighter magnitudes than TriAnd.
The spectra show that, like the TriAnd clouds, this structure has a
coherent velocity structure with low dispersion and a steady gradient with
longitude, $l$, confirming the genuine association of its members \cite{li17};
see Figure~\ref{fig:ting_vel}, orange squares.

Low-resolution spectroscopy has also been obtained for a sample of M giant stars
that span~$\approx $100$^\circ$ of the Mon/GASS structure \cite{crane03}.
The candidate Mon/GASS member M giants also show a clear trend in GSR velocity
with Galactic longitude, and appear to form a coherent sequence with both A13
and TriAnd; see Figure~\ref{fig:ting_vel}, yellow circles.

As mentioned above, Figure~\ref{fig:ting_vel} (reproduced with data from
previous work, \cite{li17}), summarizes the line-of-sight (GSR) velocity trends of M
giants in each of the three low-latitude structures.
Not only do these structures all have low dispersions ($\sim$25 km/s) relative to the stellar halo ($\sim$120 km/s)
--- which suggests that the structures themselves are real---they also appear to
{\it collectively} exhibit a continuous gradient with Galactic longitude, $l$.
This suggests that the structures may also be associated with one another, as part
of a larger structure in the outer Galactic disk.

%{\it Summary: the low latitude structures each have low velocity dispersions supporting the genuine association of the stars within them; they also form velocity sequences as functions of Galactic longitude that are continuous across the structures.}

\subsection{Stellar Populations and Other Kinematic Tracers}
 \label{sec:populations}
Motivated by the observed, low-dispersion velocity distribution of the M giant
stars in TriAnd, we sought to observe other distance tracer stars in the same
region, determine their membership, and improve the distance estimates to the
structure.
We focused on and selected RR Lyrae stars in the TriAnd region from the Palomar
Transient Factory (PTF; \cite{ptf}), using a conservative distance cut to
account for uncertainties in the RR Lyrae and M giant photometric distance
estimates, $15~\kpc~<~d_\odot~<~35~\kpc$.
We obtained spectra for $\approx$1/3 of the total number of RR Lyrae in the M
giant volume considered to be associated with TriAnd and measured radial
velocities for~these~stars~\cite{pricewhelan15}.

\begin{figure}[H]
\centering
\includegraphics[width=3.5 in]{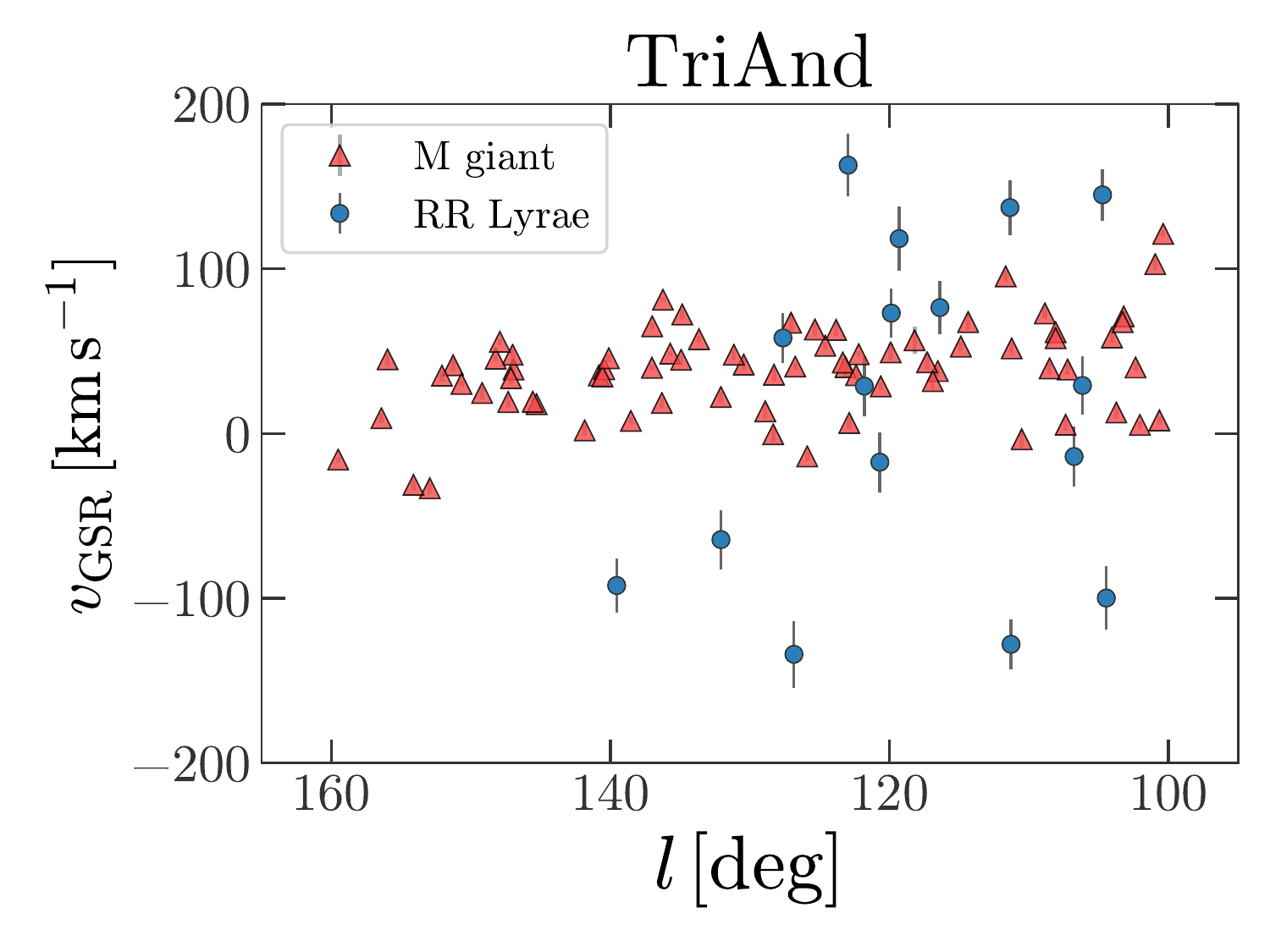}
\vspace{-12pt}
\caption{\label{fig:apw}
Comparison of the velocity distribution for M giants in the TriAnd structure
(red triangles) with velocities for RR Lyrae stars in the same region of sky
and distance range (blue circles).
Radial~velocity uncertainties of the M giant stars are typically the size of the
marker or smaller.
Uncertainties for the RR Lyrae stars are shown with gray error bars.
Note the low-dispersion sequence in the M giant velocities, unseen in the RR
Lyrae star velocities, which look like typical halo stars with a large
velocity dispersion. Reproduced from \cite{pricewhelan15}.}
\end{figure}

Figure~\ref{fig:apw} (reproduced from \cite{pricewhelan15}) shows the results
of our survey: unlike the M giants (triangles) the RR Lyrae stars (circles) show
no clear, tight velocity sequence.
By modeling both the RR Lyrae and M giants velocities as having been drawn from
a mixture of two populations---one representing a low-dispersion
sequence with varied dispersion, and one representing a halo
population with large dispersion, both Gaussian---we showed that, after
accounting for selection effects, the number ratio of RR Lyrae to M giants, \frrmg, within the overdensity is $\frrmg < 0.38$ with 95\% confidence.

%\begin{figure}[t]
%\centering
%\includegraphics[width=3 cm]{logo-mdpi}
%\caption{\label{fig:isochrones}
%Isochrones for various ages metallicities}
%\end{figure}

Since we were unable to find any RR Lyrae clearly associated with TriAnd, our
attempt to measure more accurate distances to the structure was unsuccessful.
However, the upper limit on the value of $\frrmg$ was in itself interesting.
RR Lyrae and M giants are tracers of populations with quite distinct
metallicities: stars in the horizontal branch phase of evolution are typically
only blue enough to cross the instability strip and become RR Lyrae if they have
$\feh \lesssim -1.5$; and giant stars typically only evolve to colors red enough
to become spectral class M if they have $\feh \gtrsim -1.5$.
Hence, a stellar population has to contain a significant range of metallicities
to contain both types of stars.
%In order to aid with the interpretation of this discovery, Figure~\ref{fig:isochrones} plots isochrones for two $10~{\rm Gyr}$-old populations of low and intermediate metallicity, including the color range for RR Lyrae and M spectral class.
%It emphasizes our understanding that these two types of stars are tracers of populations with quite distinct metallicities and illustrates why, while the initial aim of our survey to find more accurate distances to TriAnd 1 and 2 had failed, our results could be even more intriguing.
The~metallicity distributions in nearly all existing satellite galaxies
orbiting the Milky Way (e.g., \cite{kirby11}) are typically biased towards
low metallicities (i.e., [Fe/H] < $-$1) , such that they contain no or few M giant stars (i.e., $\frrmg = \infty$).
The largest satellites (the Large Magellanic Cloud and Sgr) are exceptions as they contain substantial metal rich (i.e., [Fe/H] > $-$1) populations; these very metal-enriched dwarf galaxies have $\frrmg \sim 0.5$ in their still-bound stellar populations \cite{pricewhelan15}.
In contrast, the local Galactic disk is an overall metal-rich population and thus has very few RR Lyrae (i.e., $\frrmg \sim 0$; \cite {amrose01}), more consistent with our findings for the stellar population of TriAnd.
Moreover, results from the APOGEE survey show the outer disk populations have [Fe/H] > $-$1 \cite{hayden15}, and hence it is also likely contain a very low fraction of RR Lyrae.

Our work on the stellar populations of the TriAnd region motivated us to look at possible associations of RR Lyrae with the M giant sequences found in Mon/GASS and A13.
For the surveyed regions of these structures, we find \frrmg\ values similar to those observed for TriAnd, and therefore consistent with membership of the Galactic disk (Sheffield et al., in prep.).

%{\it Summary: the stellar populations in these structures look like each other; they are more consistent with those in the Galactic disk rather than those observed in the stellar halo or Galactic  satellites.}

\subsection{High-Resolution Spectroscopy: Abundance Patterns}

\label{sec:abundances}

The origin of stellar associations can also be explored through measurements of
the detailed chemical abundances of their constituent stars.
It is intriguing that our finding of the low-latitude structures all having stellar
populations (as indicated by \frrmg) that look more like the disk than known
Galactic satellites (Section~\ref{sec:populations}) appears to be at odds with
prior work on abundance patterns of stars in these structures.
High-resolution spectra of 21 M giants in Mon/GASS \cite{chou2010b} showed
[Ti/Fe] lower by up to 0.4 dex compared to the mean trends known for
main-sequence stars of the Galactic disk (e.g., \cite{reddy03,bensby2014}) and
a mean offset for [Y/Fe] of about 0.2 dex at ${\rm [Fe/H]} \approx -0.5$.
A comparison with similar results for stars in Sgr
\cite{chou2010a} suggested that Mon/GASS may be more similar to Sgr than the
disk, and therefore proposed an external origin for this structure. Subsequent
work comes to similar conclusions for TriAnd stars \cite{chou11}.
%It should be noted, however, that their results do not agree with the abundances of Ti and, in particular, La, measured in the Canis Major overdensity by \cite{sbordone2005}. Their analysis was based on spectra covering a limited spectral region in the near-IR, including 11 neutral iron and 2 neutral titanium lines. Also \cite{chou2010b} caution that the effects of non-local thermodynamic equilibrium (non-LTE) may be significant, since their measurement of metallicity and Ti abundances relies on the LTE analysis of lines of neutral species,

\begin{figure}[H]
\centering
\includegraphics[width=6.2 in]{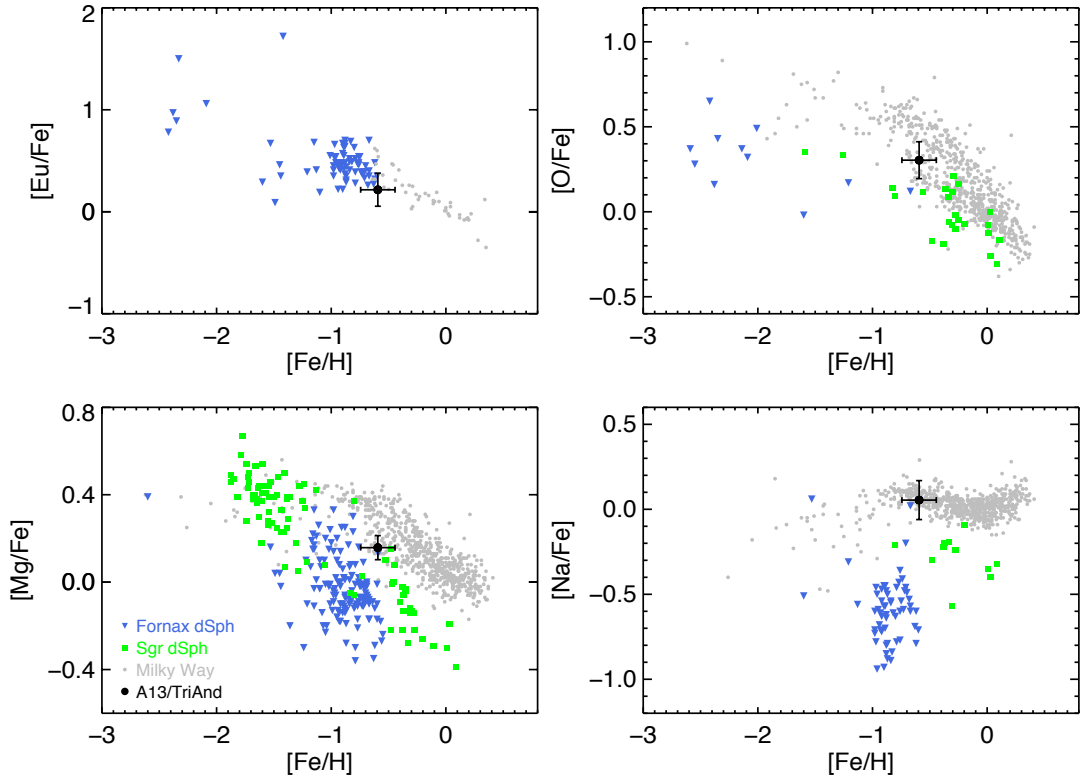}
\vspace{-35pt}
\caption{\label{fig:abundances}
Abundance patterns for the thin and thick disk (grey symbols, from  \cite{bensby2014,battistini16}), the Fornax (blue symbols, from \cite{shetrone03,letarte06,letarte10,kirby11,larsen12}) and Sgr (green symbols, from \cite{mcwilliam05,sbordone07,mucciarelli17}) dwarf spheroidal galaxies and our own results (black circle). The error bar on our measurement is the sum of the standard deviation of the stellar  sample and a systematic error of the abundance measurement.}
\end{figure}

% ALLY COMMENT: Was this the actual motivation, if we look back at the
% proposal? We were asking for time to explore the possibility of TriAnd being
% a kicked out disk population
% KVJ RESPONSE: that was because we had (embarrassing) forgotten about the Chou et al work? Don't think we need to admit that here.....
To explore the apparent contradiction between the stellar populations
and abundance work, we~have recently obtained high-resolution ($R \sim
30,000$--$47,000$) and high signal-to-noise (>$200$ per \AA) spectra of fifteen stars in the
TriAnd and A13 overdensities.
Fourteen stars were observed with the HIRES-S spectrograph at the Keck-1 telescope
\cite{vogt1994} and one star was observed using the UVES spectrograph at the
VLT (Program ID: 097.B-0770A).
%The spectral resolving power R of the HIRES spectra is 36\,000 and the UVES data have R $\sim$ 47\,000. All Keck spectra cover the full optical region, from $4800$ to $8770$ $\AA$, although the exact wavelength coverage varies as several slightly different instrument configurations were tried in an effort to get all the key lines into a single exposure. The signal-to-noise (SNR)/\AA of the HIRES spectra near 5200 \AA in the continuum at the center of the echelle order exceeds 200. For the UVES spectrum, the SNR of 50 near 5500 \AA was achieved. The fundamental parameters and chemical abundances of stars were determined using 2MASS and APASS photometry and the high-resolution spectra. All 15 stars are M-type giants with effective temperatures Teff ~ 3800 K and surface gravities log(g) ~ 1 dex. The $T_{\rm eff}$ estimates were derived using the Infrared Flux Method (Casagrande et al. 2010, 2014). Chemical abundances were measured for 6 chemical elements, including O, Na, Mg, Ca, Fe, Eu using LTE and the standard MARCS stellar model atmospheres (Gustafsson et al. 2008). We have also performed detailed analysis of the effects of non-local thermodynamic equilibrium (non-LTE) for the elements, where detailed calculations are available \cite{bergemann2011, bergemann2012, bergemann2016}, however, the NLTE corrections for the chosen lines are minor and do not change our conclusions. In what follows, we use our LTE results, because all studies of chemical abundances in dSph systems and most studies in the Milky Way to date have employed LTE with 1D hydrostatic models
Figure \ref{fig:abundances} shows the results of our preliminary analysis, with the average over all the stars in our sample presented as a black point.
We find that the stars in these structures have a very narrow metallicity spread, with a value that is consistent with the prior metallicity estimates in TriAnd stars \cite{chou11}.
The TriAnd and A13 stars also have extremely similar chemical abundances to each other, with the abundance dispersion across the combined set of stars from both structures of $\leq$$0.06$ dex for most chemical elements.
The abundances of all measured $\alpha$-elements are uniformly enhanced at a level that is consistent with the abundances of the Milky Way disk stars (gray points;  \cite{fuhrmann2004,bergemann2014,bensby2014}).
The [Mg/Fe] value is also consistent with the measurements of $\alpha$-elements for the disk stars that have been found towards the outer disk in the APOGEE survey \cite{hayden15} (with the caveat that at a given [Fe/H] there is typically an offset of $\sim$0.2 dex between [Mg/Fe] measured in the infrared with APOGEE and other optical studies).
%the stars in these structures have extremely similar chemical abundances, with the abundance dispersion across the combined set of stars from both structures $\leq 0.06$ dex  for most chemical elements.
%The TriAnd and A13 stars also have a very narrow metallicity spread, ${\rm [Fe/H]} = -0.57 \pm 0.08$ dex.
%The abundances of all measured alpha-elements are uniformly enhanced at the level of [O/Fe] = $0.48 \pm 0.07$ dex, [Mg/Fe] = $0.17 \pm 0.03$ dex, [Na/Fe] = $0.29 \pm 0.02$ dex, and [Eu/Fe] = $0.19 \pm 0.06$ dex.
%These results are similar to the average in prior abundance work for [Fe/H] \cite{chou11}.
%However, with these enhancements in other elements, the TriAnd and A13 stars are consistent with the abundances of
% the in-situ formed
%the Milky Way thick disk stars\cite{fuhrmann2004,bergemann2014,bensby2014}.
%In particular, the [Mg/Fe] value is also consistent with the measurements $\alpha$ elements for the few thick disk stars that have been found towards the {\it outer} disk in the APOGEE survey \cite{hayden15}.
%The only exception is [Ca/Fe], which is consistent with solar with a dispersion of 0.06 dex; however, the non-local thermodynamic equilibrium (NLTE) abundance correction for the \ion{Ca}{I} lines for M giants is 0.1 dex, which will bring the results in agreement with the MW disk studies that are based on FGK dwarfs, for which the NLTE correction is much smaller \cite{Merle2011}.
The abundance ratios we derive are too high for the chemical abundance patterns observed in the stars of the Galactic satellites (colored points; \cite{Bonifacio:2000,shetrone2001,shetrone03,Tolstoy:2009,deBoer:2014}), which are known to have low, typically solar ([O/Fe]) or even sub-solar ([Mg/Fe],[Na/Fe]), ratios at [Fe/H]~$\sim$$-$0.$5$.
However, note that very recent APOGEE results suggest elevated abundance levels for some elements are present in some stars in the Sgr dwarf at [Fe/H]~$\lesssim$$-$0.$4$,~so further work in this area is definitely warranted (Hasselquist et al., in prep.).

\vspace{-20pt}

%For example, Fornax, a dSph galaxy most close to TriAnd 1 and A13 in metallicity, has [Na/Fe] $\sim -0.6$ dex and [Mg/Fe] $\sim 0$ at [Fe/H] $=-1$ (Letarte et al. 2010, Kirby 2010, Lemasle et al. 2014) \footnote{Note that to avoid systematic differences between abundance measurements obtained from spectra taken with different instruments, we have chosen here to compare with the data taken with the same instrument; here UVES and FLAMES at the VLT, ESO, or with HIRES at Keck.}.

%Our results for TriAnd and A13 do not confirm the conclusions by \cite{chou2010b}. Although there could be several astrophysical reasons for the differences, this could be caused by the
We are currently exploring one explanation for the disagreement between our and prior abundance work for
stars in the same structures---that the differences in interpretation are due to differences in the observed datasets and
spectroscopic modeling techniques.
%Our observed data cover the full optical range from the near-UV to the near-IR, whereas
Past work \cite{chou2010b,chou11} analyzed a small wavelength region in the near-IR, limited to 150 \AA\ from 7440 to 7590 \AA. Because of this limitation, they could include only 11 \ion{Fe}{I} and 2 \ion{Ti}{I} lines in the determination of metallicities, and abundances. Our~full analysis, through a much wider wavelength coverage and high SNR attained for the observed spectra,
will allow us to include more then a hundred of Fe I and Fe II lines, as well as tens of Ti lines from both ionization stages (Ti II is important to check the influence of NLTE effects in Ti I). It~has~been~shown that Ti I should be not be used in abundance studies because it is very sensitive to NLTE effects \cite{bergemann2011} (see also discussion in \cite{sheffield12}).
%will allow us to include 70 lines of \ion{Fe}{I} and \ion{Fe}{II} \cite{bergemann2012}, as well as 10 lines of \ion{Ti}{II}. Moreover, it has been shown that \ion{Ti}{I} should be not be used in abundance studies because it is very sensitive to NLTE effects \cite{bergemann2011} (but see also discussion in \cite{sheffield12}).
We are investigating the sensitivity of abundance diagnostics to the line selection and wavelength regimes by performing test computations on our own data, using a reduced line-list and comparing our [Fe/H] and [Ti/Fe] results between this and the analysis using the full line-list.
%In order to explore the sensitivity of abundance diagnostics to the line selection and wavelength regimes we performed preliminary test computations on two stars in our own data, using a reduced line-list. We have found that using the line-list from \cite{chou2010a}, the metallicities are over-estimated by $\sim 0.25$ dex and [Ti/Fe] are under-estimated by $0.4$ dex, compared to the results using the complete line-list. This suggests that the choice of the line-list and diagnostic spectral band (full optical or near-IR) as a plausible explanation  of the apparently discrepant results.
%{\it Summary: the abundance patterns  of the low-latitude structures are similar to those of the thick disk of our Galaxy.}

Overall, comparing to abundances of disk stars and satellite populations obtained using analogous~data sets and reduction techniques, our current results indicate, in contrast to the prior~work~\cite{chou2010b,chou11}, that the birthplace of stars in TriAnd and A13 could be the outer Galactic~disk~(Bergemann~et al.,~in~prep.). The differences in our conclusions provide powerful motivation for a uniform survey of abundances for a much larger number of stars towards the Galactic anticenter and extending tens of kpc from the Sun (e.g., \cite {hayden15}), as well as at comparable metallicities in the few satellite galaxies large enough to host such metal-rich populations.

%it is unlikely that the TriAnd 1 and A13 stars originate from a disrupted dSph galaxy

\subsection{Mapping Main-Sequence Stars in the Low-Latitude Structures}

While we have concentrated on follow-up studies of the known low-latitude structures as traced by M giants selected from 2MASS, knowledge of the spatial distribution of MSTO stars towards the anticenter region has been further refined using photometry from the PanSTARRS1 survey \cite{slater14,Morganson:2016}.
Recent  studies using the
 SDSS \cite{xu15} and Pan-STARRS1 (Lurie et al., in prep.) surveys 
 employ the technique of subtracting color-magnitude diagrams (CMD's) derived from fields in their photometric data which were symmetrically placed at equal and opposite Galactic latitudes and at the same Galactic longitudes, analogous to that used for M giants in the original discovery paper for the TriAnd Clouds~\cite{rochapinto04}.
The vast number of MSTO stars allowed denser regions, closer to the Galactic plane and at smaller heliocentric distances  to be explored.
Both MSTO studies show overdense and clearly distinct  arcs of stars oscillating between the Northern and Southern Hemispheres as the heliocentric distance was increased towards the anticenter of our Galaxy.
%The vast numbers of stars in these surveys allowed more clear identification of these global structures as clearly distinct, separate features.

%{\it Summary: the overdensities around the outer Galaxy, oscillating between the northern and southern hemispheres have been traced to smaller-scale oscillations all the way to the Solar Neighborhood.}

\subsection{Connecting the Low-Latitude Structures to Velocity Structure Near the Solar Radius}

% APW COMMENT: what exactly do we mean by "local"?
Coincident with studies exploring structures at the very outer edge of our
Galactic disk, large scale spectroscopic surveys have allowed a detailed
re-examination of the {\it local}  distribution of stellar velocities, within a few kpc of the Sun.
Using data from the SDSS \cite{widrow12,yanny13} and the RAdial Velocity
Experiment (RAVE; \cite{rave,williams13}), asymmetries between the Northern and
Southern Galactic Hemispheres have been seen in the density and velocity
distributions of stars in the vicinity of the sun.
Looking~$\sim$2~kpc~out towards the Galactic anticenter, the Large Sky Area Multi-Object Fiber Spectroscopic Telescope (LAMOST; \cite{cui12,deng12,zhao12}) finds similar asymmetries in radial and vertical velocities~\cite{carlin13}.
The scale and sense of these asymmetries indicate moderate systematic motions (of order a few km/s) of stars within the disk perpendicular to the plane, suggesting both vertical movement of the midplane, and compression and rarefaction of the vertical scale (referred to as ``bending'' and ``breathing'' modes respectively---see, e.g., \cite{widrow14}).
It is natural to associate these asymmetries in motion as a local manifestations of the oscillations traced in space over much larger scales \cite{xu15,pricewhelan15}.
%{\it Summary: small-scale, systematic vertical motions of and within the disk have been detected in the Solar Neighborhood.}

\section{The Nature of Structures Around the Outer Disk---Summary of Theoretical Interpretations}
\label{sec:theory}

% APW COMMENT: I think Chervin's simulation may be fine as our "cartoon" ... may
% be that the structures are too complex to simplify into a 1D cartoon

% \begin{figure}[t]
% \centering
% \includegraphics[width=4 in]{cartoon.pdf}
% \caption{\label{fig:cartoon}
% TODO: Cartoon of oscillations in space and velocity}
% \end{figure}

% TING COMMENT: everything above line 251 is really summary of observation.
% Maybe this should go to Section 2.6? It is kinda confusing that this section
% is for theory but it were all about observations for the first 20 lines
% Figure \ref{fig:cartoon} is a cartoon which summarizes all the observational works and their implications, by showing the approximate locations and amplitudes of the spatial and velocity structures that have been identified. (Note that the figure is misleading as the structures are {\it not} consistent morphologically with concentric rings that are axisymmetric about the Galactic center.)

The observations summarized in Section \ref{sec:obs} indicate that:
\begin{itemize}
\item the low-latitude structures---Mon/GASS, TriAnd, and A13 ---each have
      low velocity dispersions supporting the genuine association of the
      candidate member stars;
\item Mon/GASS, TriAnd, and A13 share a continuous sequence in mean GSR velocity
      as a function of Galactic longitude, suggestive of associations between
      these structures;
\item the stellar populations in the structures (as indicated by \frrmg) are all
      more consistent with those in the Galactic disk rather than those observed
      in the stellar halo or Galactic satellites;
\item the abundance patterns of stars in TriAnd and A13 are similar to those
      found in the disk of our Galaxy (although the discrepancy with prior work in this conclusion is still under investigation);
\item the low-latitude structures (around the outer disk) may be connected to
      oscillating density and velocity structure on smaller scales, traced all
      the way back to the Solar neighborhood;
\end{itemize}

Taken together, we conclude that: (i) there is mounting evidence that
Mon/GASS, TriAnd, and A13 represent populations of stars formed in the disk that
now exist at extreme radii and heights above the Galactic disk;
(ii) these structures are likely associated and part of a global system of vertical
disk oscillations that can be traced all the way to the velocity asymmetries
seen in the solar neighborhood; and
(iii) the stellar populations in
%(and, in some work, chemical abundances of)
these structures are
inconsistent with a picture in which they formed in a dwarf galaxy.

One natural interpretation of these collected observations is that the oscillations represent the response of the disk to an external perturbation---for example, the impact of a satellite galaxy that has been  transmitted and amplified by its wake in the dark matter halo \cite[][]{weinberg06,gomez16}.
Prior work has already pointed to this as a possible explanation for the existence of Mon/GASS \cite{kazantzidis08,younger08}, with Sgr being pointed to as a plausible culprit for the perturbation \cite{purcell11}.
It has also been demonstrated how perturbations from a satellite on an orbit perpendicular to the disk could lead to bending (at low relative impact velocity) and breathing (at higher relative velocity) modes that would be observed in the solar neighborhood as asymmetries in the local velocity distribution  \cite{widrow14} and on larger scales as rings \cite{donghia16}.
(Note that breathing modes can also be induced by non-axisymmetric features in the disk such as the bar and spiral arms; \cite{monari16}.)
Simulations have also shown that Sgr could be responsible for local velocity structure  \cite{gomez13}.
Such interactions and corresponding disk features have been found to naturally occur in cosmological simulations \cite{gomez16}.

\begin{figure}[H]
\centering
\includegraphics[width=5 in]{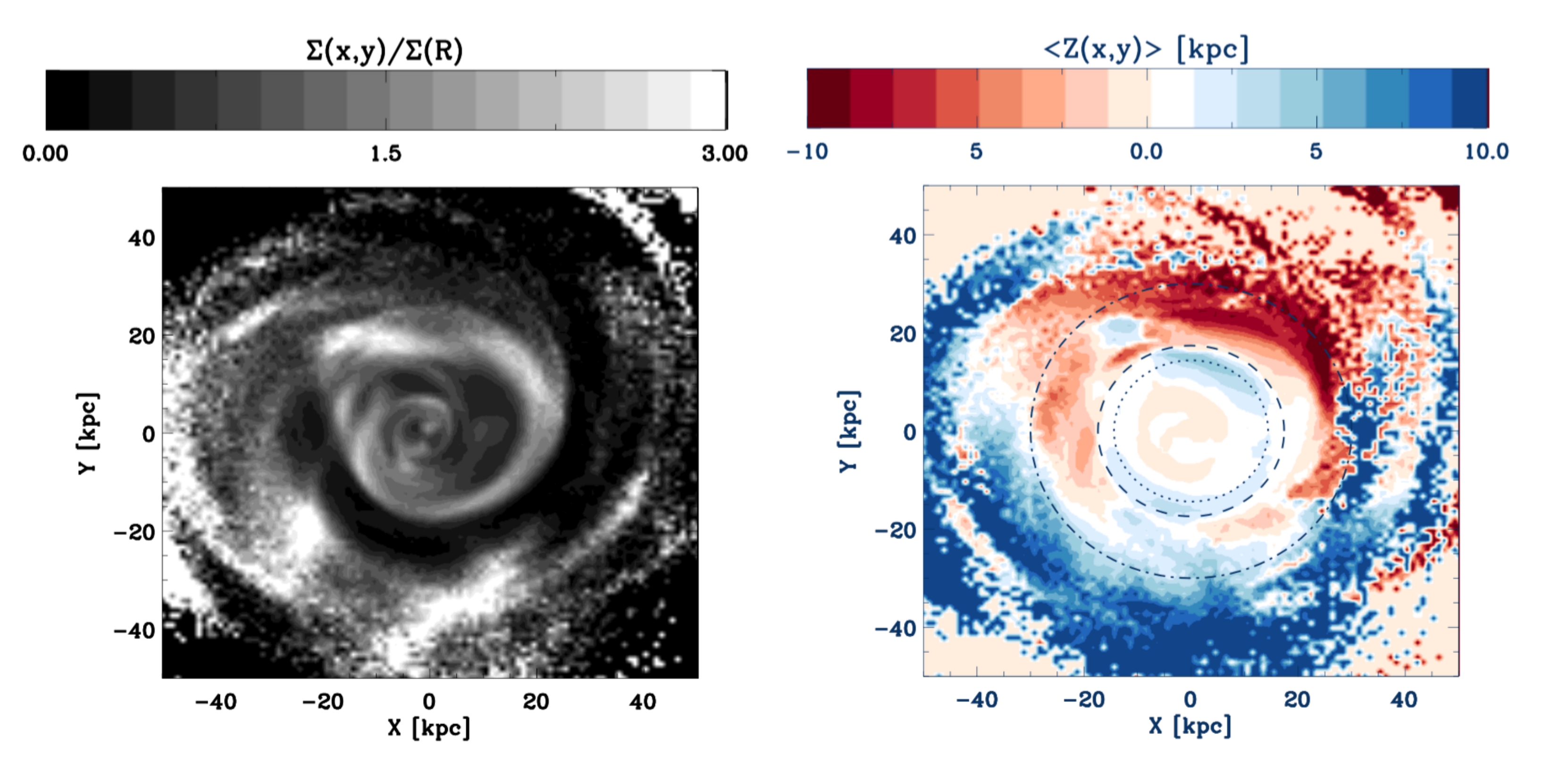}
\vspace{-12pt}
\caption{\label{fig:chervin}
% APW COMMENT: needs a caption - waiting until we have final version of figure.
Visualizations at the end-point of a simulation of the impact of the Sgr dwarf galaxy on the Milky Way's disk from work currently in preparation (Laporte et al.).
The simulation encompassed the last $\sim$6 Gyrs of the interaction.
Sgr had encountered the disk 6 times during this time and had a final mass of $\sim 3\times 10^9$ $M_\circ$.
The left hand panel shows density fluctuations around the mean at each annulus in the disk.
The right hand panel shows the average positions of particles above/below the disk at each point.
Note that material is kicked to as much as 10 kpc from the midplane at distances of 20--30 kpc from the Galactic center.}
\end{figure}

Figure \ref{fig:chervin} illustrates these ideas with the results of simulations from our own recent work.
Using~simulations~of a disk disturbed (separately) by satellites on orbits that
mimic those expected for the Large Magellanic Cloud and Sgr \cite{laporte16}, we
extend the prior theoretical backdrop that looked at Mon/GASS to examine whether
the extreme ($R,Z$) locations of TriAnd stars could fit within the same picture.
With different masses and orbits (and consequently different interaction
strength, timings, and durations) these satellites necessarily induce distinct
but overlapping signatures on the global structure of the disk.
In more recent work, we found a model that was capable of reproducing the scales of the observed disturbances (radial wavelength and amplitude in space, as well as magnitude of offsets in velocity locally).
The model required:
the interaction of Sgr with the disk of the Milky Way to be followed for several passages longer than prior work
(note that the length of Sgr's streams indicate that it has been impacting the disk for several pericentric passages e.g., \cite{law10});
Sgr  to have sufficient initial mass and density to impact the disk in the last Gigayear with a remaining mass of $\sim$3$\times10^{9}$~$\msun$;
and the disk to be realized with stars existing as far out as $40~\kpc$ from the Galactic center to populate the regions corresponding to TriAnd.
The interaction with the LMC modified the overall morphology of the structures induced, but was not sufficient alone to explain their properties.
The full details of these results will be discussed in an upcoming paper (Laporte et al., in prep.).

\vspace{-6pt}

\section{Discussion---Observational Prospects}
\label{sec:Discussion}
\vspace{-10pt}
\subsection{The Milky Way}

%It is interesting to place the current work in the context of ongoing and near-future studies of our Galaxy.
While the connections that have already been made between the different
low-latitude structures and the disk population are convincing, there are
several possible directions for further observations to strengthen these claims.
Kinematic and abundance information and density measurements for more tracers
and over a larger volume
would greatly facilitate an informative comparison to theoretical work, allowing
more detailed interpretations of the history of our Galaxy.

The most obvious direction is to obtain proper motions and more accurate
distance estimates to the known features, using the candidate members discussed
in this work to search for other tracers.
For example, proper motions for the ``Anti-Center Stream'' (ACS) (a thin, coherent density structure, which may
or may not be part of the larger Mon/GASS system; \cite{li12}) indicate that
stars in the ACS are not actually moving along the spatial extent of the stream \cite{carlin10}.
This is inconsistent with expectations for the behavior of debris from a
destroyed satellite.
If similar measurements of proper motions of stars in all of the low-latitude
structures showed significant motion perpendicular to the Galactic disk, this~
would~provide even more evidence of a disk origin and connection to local
oscillations (first results are just being reported in this region \cite{deboer17}).
With precise proper motions, the velocity information would also place important
constraints on dynamical models of the disk (see Section~\ref{sec:conc}).
In upcoming data-releases, astrometric measurements from the {\it Gaia} mission
\cite{gaia} are poised to provide these data.
Expected proper motion uncertainties for the closest M giant stars ($\approx
$5$~\kpc$) in Mon/GASS correspond to tangential velocity uncertainties of $\approx
$1--2$~\kms$ for M giants with tangential velocities $<$50$~\kms$.
For the farthest known M giant stars in TriAnd ($\approx 35~\kpc$), tangential
velocity uncertainties will be $\approx $7--12$~\kms$ for M giants with
tangential velocities $<$50$~\kms$.

The next decade will see first light for the Large Synoptic Survey Telescope (LSST; \cite{ivezic08}), which will survey the sky to the same depth as SDSS every three days.
Over time, these data can be combined to detect MSTO stars out to 100 kpc. Hence the exquisite maps from SDSS and PanSTARRS that exploited the dense coverage of this tracer to reveal disk oscillations out to Mon/GASS can be extended to A13 and TriAnd and beyond.

Another complimentary direction for future data is to extend spectroscopic maps to fainter magnitudes and global scales.
%{\it Gaia} will be able to tackle this with distances, proper motions, and radial velocities.
The {\it Gaia} catalogue will soon be enhanced by surveys from several ground-based, wide-field, multi-object spectrographs on larger telescopes capable of reaching to fainter magnitudes (e.g., WEAVE on the William Herschel Telescope, 4MOST and the Prime Focus Spectrograph on the Subaru telescope---\cite{weave,4most,pfs}).
However, the reach of these surveys towards lower latitudes all the way to the disk plane 
%in the inner Galaxy
will be limited by extinction.
An infrared spectroscopic survey, such as APOGEE \cite{apogee} , could~overcome this limitation.
APOGEE is capable of building catalogues of radial velocities and abundance patterns  for 15 chemical species across the Galactic disk, bulge, bar, halo, star clusters, and dwarf galaxies--- the type of homogeneous data set needed to compare the low latitude structures to other Galactic components.
%. This program is uniquely able to provide a homogenous dataset from which to compare the chemo-dynamical properties of the disk, halo, and satellite galaxies with which to compare these low latitude structures.
%APOGEE is one of only a handful of large scale spectroscopic programs still operating in the Northern hemisphere where these structures are best studied.

% APW COMMENT: you mention LSST in next section, but what about how LSST will
% provide deep photometry - map far down the main sequence for stars in these
% structures.

% APW COMMENT: I feel like we need to say more about APOGEE, future
% spectroscopic surveys like DESI, PFS.

\subsection{Other Galaxies}

% TING COMMENT: There are also a significant amount of studies on "galaxy
% morphology and warps" with distance galaxies. For example:
% http://adsabs.harvard.edu/abs/2016JKAS...49..239A
% You may want to mention these studies (reference therein) as well....

% APW COMMENT: this sentence comes a bit out of the blue -- needs a better
% introductory or transitional statement
%The great advantage of star-count studies is the ability to reach extremely low surface brightnesses. For example, TriAnd is estimated to have a surface brightness lower than $\Sigma <$32 mag/arcsec$^2$ \cite{majewski04}.
%For example, TriAnd I contains ?? M-giant stars spread over an area on the sky of roughly ????, so ? giants/deg$^2$.
%Adopting a 10 Gyr-old isochrone for this population \cite[from the Padova structure][]{}, each M-giant has an associated total stellar luminosity of ??? in the ?? band. Hence, the equivalanet urfc ebright ness would be .....

% APW COMMENT: this section needs some work, but ran out of steam...

TriAnd is estimated to have a surface
brightness below 32 mag/arcsec$^2$ \cite{majewski04}, so studying similar features around other galaxies presents a significant challenge.
Nevertheless, there are growing samples of galaxies within and beyond the Local Group being mapped to depths close to  this target.

For nearby galaxies, these levels are approached through star counts studies, most spectacularly for the case of our nearest neighbor, the Andromeda Galaxy, where giant star counts have revealed a significantly extended and richly featured outer stellar disk \cite{ferguson02,ibata05}.
Analogous studies have been carried out for galaxies up to distances of several Mpc (e.g., \cite{monachesi13,crnojevi16}), although the focus of these studies has typically been on detecting the stellar halos of these objects.
The great advantage of star-count studies is the ability to reach extremely low
surface brightnesses---depths below 30 mag/arcsec$^2$
have been estimated for the density profile of M31's and other galaxies' stellar halos
% Is the bold necessary?
\cite{ibata07,harmsen17}.

Star count studies are limited to the galaxies in the volume within which stars can be resolved.
Several dedicated surveys have made innovative advances in studying galaxies to low surface-brightness using a variety of techniques to reach limits below 30 mag/arcsec$^2$ in integrated light (e.g., \citep{delgado10,vandokkum14,duc15,spavone17}), although these studies face the twin challenges of  calibration and Galactic cirrus to overcome.

Looking to the future,
NASA's proposed Wide Field InfraRed Survey Telescope (WFIRST), with its wide field of view and high resolution, offers the possibility of extending the deep star-count sensitivities now achieved in MW and M31 to all galaxies within 10 Mpc \cite{spergel13}.
In  the next decade, images in integrated light from LSST can be combined to be sensitive to slightly shallower depth ($\sim$29~mag/arcsec$^2$,~see \citep{ivezic08}), but for vast numbers (many millions!) of galaxies.
% TING COMMENT: about 29 mag/arcsec - I tried to find the number "29
% mag/arcsec^2" in the reference but I could not. I doubt this number is true
% if you are talking about stellar resolved overdensities. TriAnd is 32 mag/
% arcsec^2 and can be discovered with a much smaller telescope. Maybe you are
% talking about galaxies (star not resolved) only... But it is kinda confusing
% KVJ RES{PONSE: I think people are working on this and want to leave it in as aspirational. - its buried here or in the scinece book. The point about LSST is the vast number of galaxies.

\section{Conclusions---What Might These Structures Tell Us about Galaxies?}

%{\it KVJ---Reminder to self to add these somewhere} \\
%arXiv:1706.01900 ---Title: Milky Way Tomography with K and M Dwarf Stars: the Vertical Structure of the Galactic Disk \\
 %D'Onghia et al 2016 on sat and disk interaction \\
% Bovy et al 2015 - power spectrum of vel in disk \\
% Kazantzidis08 - rings etc \\
% schwarzkopf 01 \\
 %Zarik=tsky 97 - lopsided gals and accreiton

\label{sec:conc}

The above sections summarize observational evidence for large scale vertical
oscillations of the Galactic plane present in the Solar neighborhood and
reaching out beyond the traditional edge of our stellar disk.
We have also discussed theoretical studies that suggest that these oscillations
could be caused by, and contain the signatures of, ongoing interactions of the
Milky Way with its satellite system.
Moreover, there are numerous observational prospects for extending this work
both to map the Milky Way globally and to look for analogous features around
many other galaxies.

Now that we have a physical picture of the origin of such planar undulations, as well as prospects for mapping them further within the Milky Way and detecting analogous substructures around other galaxies, we can move on to discussing how useful they are for constraining the dynamics and evolution of galaxies.
While the mere existence of these low surface brightness structures around the outskirts of galaxies is interesting, they  contain only a tiny fraction of the stars in galaxies and these are spread out over a large area---these properties naturally make such structures difficult to map, either because their unique signatures can be lost in the foreground star counts (e.g., in the Milky Way), or~because the required surface brightness limits for detection are prohibitively low (for~integrated~light).

Conversely, these features may prove to be particularly powerful probes of interactions and histories, precisely because they contain so little mass: they can be modeled as test particles responding to an external perturbation.

Below are just three examples of where these structures could promise new insights into some classic questions in galactic astronomy.
\begin{itemize}
\item{\it Disk heating mechanisms ---}
It has been understood for a long time that disks can evolve significantly due to mergers, major or minor, and hence that their current structures bear witness to their accretion history \cite{toth92,quinn93,walker96,velazquez99}.
This understanding has fueled a significant literature on simulations investigating the importance of the heating of galactic disks in response to encounters with other dark matter halos (that may or may not contain their own galaxies) \cite{font01,ardi03,benson04,stewart08,hopkins08,villalobos08,purcell09,kazantzidis09,sachdeva16,moetazedian16}.
In general, these studies have concentrated on the overall effects of many encounters on global properties, such as the thickness and vertical velocity dispersion in disks.
The results of these simulations have traditionally been compared to the spatial and velocity scales in samples of galaxies.
In contrast, the identification and mapping of vertical waves associated with ongoing interactions in the Milky Way gives us the opportunity to dissect individual disk heating events in progress (e.g., the impact of Sgr).
We can use this to check our understanding of the mechanism directly and in detail rather than assessing its importance through collective effects and longterm, phase-mixed signatures.
\item{\it Stellar halo formation processes---}
%The last decade has seen increasing interest in assessing how much of the content of stellar halos could be made from stars originally formed in the disks of the galaxies that they surround. Hydrodynamical simulations of galaxy formation suggest that tens of percent of the stars in the inner halo might be formed this way\cite{abadi06,zolotov09,zolotov10,font11,mccarthy12,tissera13,tissera14,pillepich15,cooper15}.
The last decade has seen increasing interest in assessing how much of the content of our stellar halo could be made ``in situ''
along with other components of the Milky Way rather than accreted from other objects. Hydrodynamical simulations of galaxy formation suggest that tens of percent of the stars in the inner halo might be formed ``in~situ'', either along their current orbits or ``kicked-out'' from orbits that were originally in the disk \cite{abadi06,zolotov09,zolotov10,font11,mccarthy12,tissera13,tissera14,pillepich15,cooper15}.
Preliminary arguments for the existence of an ``in situ'' population were based on transitions in the density or orbital structures of stellar halos (e.g., \cite{carollo07}).
However, such transitions were also found to occur naturally in purely-accreted models of stellar halos \cite{deason13}.
More convincing observational evidence for stars in the halo that have been ``kicked-out'' of the disk  is just beginning to emerge through studies that look for stars with halo-like kinematics, but disk-like abundances around M31 \cite{dorman13} and the Milky Way \cite{sheffield12,hawkins15,bonaca17}.
The work outlined above in Sections ~\ref{sec:obs} and~\ref{sec:theory} adds new perspectives on the ``kicked-out-disk''  stellar halo formation process with the detection and modeling of disk stars that may be in transition from the disk to the halo.
\item{\it Galactoseismic probes of interactions and dark matter ---}
The response of a disk to an encounter will depend on its own properties, the properties of the dark matter halo in which it is embedded, and the mass and orbit of the perturbing satellite.
This leads to the suggestion that, analogous to helioseismic investigations of the structure of our Sun, maps of a disk response---such as those described in Section \ref{sec:obs} for our Milky Way---might be similarly inverted to tell us about, for example,  the structure of the dark matter halo
\cite{widrow12}.
Indeed, recent investigations into the signatures of encounters in the very outskirts of extended HI disks have successfully used simulations combined with an analytic understanding to find how the observed characteristics of the disturbed gas can be simply related to properties of the perturbing object \cite{chakrabarti09,chakrabarti11b,chang11}.
\end{itemize}

Overall, the breadth and depth of studies summarized in this \documentname\ are indicative of the developing interest in the new discipline of Galactoseismology, fueled by the imminence of another leap in the scale of available data sets.
The remaining challenge is to build the models and appropriate tools to compare to obervations that can take full advantage of these opportunities to significantly advance our understanding of galactic dynamics and evolution.

%Overall, the breadth and depth of studies summarized in this \documentname\ are indicative of a new discipline emerging at the right time to take advantage of the onslaught of huge data sets, which will motivate and inform detailed models and significantly advance our understanding of galactic dynamics and evolution.

%%%%%%%%%%%%%%%%%%%%%%%%%%%%%%%%%%%%%%%%%%
\vspace{6pt}

%%%%%%%%%%%%%%%%%%%%%%%%%%%%%%%%%%%%%%%%%%
%% optional
%\supplementary{The following are available online at www.mdpi.com/link, Figure S1: title, Table S1: title, Video S1: title.}

%%%%%%%%%%%%%%%%%%%%%%%%%%%%%%%%%%%%%%%%%%
\acknowledgments{Much of the work reviewed in this paper was made possible by NSF grant AST-1312196.
K.V.J. was supported by NSF grant AST-1614743 while writing the review.}

%%%%%%%%%%%%%%%%%%%%%%%%%%%%%%%%%%%%%%%%%%
\authorcontributions{This paper summarizes results from the team of listed authors. It was written by K.V.J. and A.M.P.-W., with section contributions from M.B. Figures were made by A.M.P.-W. (using data consolidated by T.S.L.)  and C.L. All the authors reviewed and commented on the drafts.}

%%%%%%%%%%%%%%%%%%%%%%%%%%%%%%%%%%%%%%%%%%
\conflictsofinterest{The authors declare no conflict of interest.}

\end{document}